\documentstyle[preprint,aps]{revtex}
\tighten
\draft
\widetext
\input epsf
\preprint{HUTP-99/A039, NUB 3203}
\bigskip
\bigskip
\begin{document}
\title{TeV-scale Supersymmetric Standard Model with\\ 
Higgs as a Slepton}
\medskip

\author{Zurab Kakushadze\footnote{E-mail: 
zurab@string.harvard.edu. Address after September 1, 1999: C.N. Yang 
Institute for
Theoretical Physics, State University of New York, Stony Brook, NY 11794.}}

\bigskip
\address{
Jefferson Laboratory of Physics, Harvard University,
Cambridge,  MA 02138\\
and\\
Department of Physics, Northeastern University, Boston, MA 02115}
\date{August 2, 1999}
\bigskip
\medskip
\maketitle

\begin{abstract}
{}Recently it was pointed out that in the TeV-scale brane world there is a
logical possibility where the electroweak Higgs can be identified with a
fourth generation slepton. In this paper we address various issues in this
four-generation TeV-scale Supersymmetric Standard Model with Higgs as a
slepton. In particular we discuss how to achieve proton stability by
suppressing the corresponding  baryon number violating operators via
gauging (a discrete subgroup of) the baryon number $U(1)$
symmetry. Dimension five lepton number violating operators which would
result in unacceptably large neutrino masses can be similarly suppressed
via gauging a discrete subgroup of the lepton number $U(1)$ 
symmetry. In fact, the
four generation feature allows for a novel higher dimensional mechanism for
generating small Majorana neutrino masses. We also discuss how to achieve
gauge coupling unification, which can be as precise at one loop as in the 
MSSM, and point out a possible geometric embedding of the corresponding
matter content in the brane world
context. Finally, we discuss adequate suppression of flavor changing
neutral currents in this model, and also point out a novel possibility for
supersymmetry breaking via a non-zero F-term of the fourth generation
lepton superfield.    
\end{abstract}

\section{Introduction}

{}The electroweak symmetry is believed to be broken via the Higgs
mechanism. In the non-supersymmetric Standard Model one can introduce a
complex scalar field that can play the role of the Higgs. In the
supersymmetric extensions of the Standard Model, however, adding a single
chiral Higgs superfield is not possible due to the corresponding gauge 
anomaly. 

{}If one does not wish to introduce superfields with exotic quantum
numbers, then the anomaly cancellation allows only two possibilities. We
can introduce either ``vector-like'' Higgs pairs or complete
generations. The former possibility has been explored in detail, and the
supersymmetric extension of the Standard Model with one ``vector-like''
Higgs pair is known as the Minimal Supersymmetric Standard Model
(MSSM). On the other hand, identifying the Higgs as a part of a complete
generation is also a logical possibility. This possibility stems from the
fact that the $SU(3)_c\otimes SU(2)_w \otimes U(1)_Y$ quantum numbers of
the electroweak Higgs are the same as those of an $SU(2)_w$ doublet lepton.
Thus, one can attempt to identify the electroweak Higgs with the scalar 
superpartner of the electroweak doublet lepton (that is, the corresponding
slepton) in one of the generations.   

{}It is, however, non-trivial to realize this Higgs as slepton
scenario. First, within the usual paradigm where the fundamental Planck
scale $M_P\sim 10^{19}$ GeV this is impossible due to the well known rules for
writing supersymmetric Lagrangians. Thus, in such a scenario
the up-quark masses would have to be
generated via the couplings of the form
$L^* Q U$, which are suppressed by the factor
$M_{\rm SUSY}^2/M_P^2\sim 10^{-32}$. 

{}However, the situation is rather different if the fundamental Planck scale
$M_{Pf}$ 
is around a TeV \cite{TeV}. In particular, now the desired couplings of the
slepton Higgs to the Standard Model fermions can be generated via the
couplings of the form $X^* L^* Q U$ in the K\"ahler potential, where $X$ is
the ``spurion'' superfield whose F-term $\langle F_X \rangle\sim M^2_{Pf}$ 
breaks supersymmetry. Note that in this context no additional suppression
appears in the corresponding Yukawa couplings. This observation was
recently utilized in \cite{GK} where the Higgs as a slepton scenario was
proposed. 

{}There is, however, another non-trivial constraint arising from the neutrino  
masses. Thus, we cannot identify the Higgs with the superpartner of a
known lepton, which is due to the fact that, via a mixing with neutral
gauginos, the corresponding neutrino, which is the superpartner of the 
neutral Higgs, invariably acquires a see-saw type of mass of order of 10
GeV. For this reason, in \cite{GK} it was proposed that there exists the
fourth generation, and the Higgs is identified with the corresponding 
slepton.      

{}In \cite{GK} it was shown that such a scenario is not ruled out by the
present data, but at the same time can conceivably be tested by LEP-II or
the Tevatron. In particular, the fourth generation neutrino picks up a
see-saw type of mass, which one requires to be at least 45 GeV to comply
with the Z-decay data. This then implies that at least one out of three 
neutralinos in this scenario is lighter than 65 GeV. In fact, the
neutralino production cross sections appear to be testable by the
LEP-II data.   

{}Even though the above scenario is not excluded experimentally, it is
desirable to address various issues invariably arising in the context of
TeV-scale quantum gravity. Thus, the purpose of this paper is to address
within the above proposal some of these issues such as proton stability,
neutrino masses, gauge coupling unification as well as suppression of
flavor-changing neutral currents (FCNCs). These issues were discussed in
detail in the context of the three-generation TeV-scale Supersymmetric 
Standard Model (TSSM) 
proposed in \cite{TSSM}. The gauge coupling unification in
the TSSM, which is as precise at one loop as in the MSSM, was discussed in
\cite{TSSM} in the context of the higher dimensional scenario proposed in 
\cite{dienes}\footnote{TeV-scale compactifications were originally
discussed in \cite{Ant} in the context of supersymmetry
breaking. Kaluza-Klein threshold corrections to gauge couplings were
first studied in \cite{TV}.}, 
where gauge coupling unification occurs via Kaluza-Klein
(KK) thresholds. Higher loop effects were shown to be subleading in this
context due to the underlying ${\cal N}=2$ supersymmetry at the heavy KK
levels in \cite{TSSM,Taylor}. Proton stability and adequate suppression of
dimension five lepton number violating operators (which would lead to
unacceptably large Majorana masses for neutrinos) was achieved in in the
TSSM in \cite{proton} by gauging an anomaly free discrete ${\bf Z}_3\otimes 
{\bf Z}_3$ gauge symmetry. Finally, adequate suppression of FCNCs in the
TSSM was discussed in \cite{flavor}, in particular, via introducing an
anomaly free non-Abelian discrete flavor symmetry $T_L\otimes T_R$ (here
$T$ is the tetrahedral subgroup of $SU(2)$) accompanied by an Abelian
(continuous or discrete) bulk flavor symmetry.    

{}In this paper we discuss the above issues in the context of the
four-generation TeV-scale Supersymmetric Standard Model with Higgs as a
fourth-generation slepton. To distinguish it from the TSSM, we will refer
to this model as the TSSM4. As we will discuss in the remainder of this
paper, proton stability in the TSSM4 can be guaranteed by gauging an
anomaly free bulk $U(1)_{\cal B}$ symmetry related to the baryon
number. This, in fact, was originally proposed in \cite{anto}. 
However, this $U(1)_{\cal B}$ symmetry must be broken, and, if it is broken
completely, then in certain cases there might be induced $\Delta B=1$ 
dimension five
operators of the form $QQQL/M_P$ (here $M_P$ is the four dimensional Planck
scale), which are well known to be disastrous for
proton stability via a one-loop graph involving a chargino exchange
\cite{wein}. To guarantee proton stability we propose gauging a discrete
subgroup of $U(1)_{\cal B}$ very much along the lines of \cite{proton}. 
In fact, it actually suffices to gauge a discrete ${\bf
Z}_2$ subgroup of $U(1)_{\cal B}$ for this purpose. Other dangerous higher
dimensional baryon number violating operators, namely, those 
with $\Delta B=2$, are not forbidden by this ${\bf Z}_2$ discrete symmetry.   
To achieve their suppression one can gauge a ${\bf Z}_4$ (instead of ${\bf
Z}_2$) subgroup
of $U(1)_{\cal B}$. In fact, for this purpose it suffices to have the
unbroken ${\bf Z}_2$ discrete symmetry with the $U(1)_{\cal B}\rightarrow
{\bf Z}_2$ (or even ${\bf Z}_4\rightarrow{\bf Z}_2$) breaking taking place
on a distant brane.    

{}Suppression of the dangerous dimension five lepton number violating
operators in the TSSM4 is achieved by gauging a discrete ${\bf Z}_3$
subgroup of an anomaly free $U(1)_{\cal L}$ symmetry related to the lepton
number. As we point out in the following, gauging the full $U(1)_{\cal L}$ 
symmetry in the TSSM4 context is not possible for the reason that the Higgs
is identified with the corresponding fourth generation slepton, and the
desired Yukawa couplings would be absent if $U(1)_{\cal L}$ were gauged in the
bulk. Its ${\bf Z}_3$ subgroup, however, which is completely adequate for
the purposes of suppressing dimension five lepton number violating
operators, is compatible with the Yukawa couplings of the Higgs with the
Standard Model fermions.

{}The gauge coupling unification in the TSSM4 is based on the same idea as
in the TSSM, but a concrete realization is a bit more non-trivial. In
particular, we discuss possible geometric embeddings (in terms of orbifold
compactifications) of the TSSM4 (with its discrete symmetries) in the brane
world context. (For general discussions of embedding the TeV-scale quantum
gravity scenario in the brane world context, see \cite{anto,ST,BW}.)
Finally, we also discuss how to suppress FCNCs in the TSSM4 (which is very
much along the lines of \cite{flavor} with some differences due to the
presence of the fourth generation) as well as some other issues. 
  
\section{TSSM4}

{}In this section we briefly review the TSSM4 proposed in \cite{GK}. The
gauge group of this model is the same as in the MSSM (or the TSSM), that
is, $SU(3)_c\otimes SU(2)_w\otimes U(1)_Y$. The light spectrum\footnote{By 
the light spectrum we mean the states which are massless before the 
supersymmetry/electroweak symmetry breaking.} of the model is ${\cal N}=1$ 
supersymmetric, and along with the vector superfields transforming in 
the adjoint
of $SU(3)_c\otimes SU(2)_w \otimes U(1)_Y$ we also have the following 
chiral superfields:
\begin{eqnarray}
 && Q_i=4\times ({\bf 3},{\bf 2})(+1/3)~,~~~
 D_i=4\times ({\overline {\bf 3}},{\bf 1})(+2/3)~,~~~
 U_i=4\times ({\overline {\bf 3}},{\bf 1})(-4/3)~,\nonumber\\
 && L_i=4\times ({\bf 1},{\bf 2})(-1)~,~~~E_i=4\times ({\bf 1},
 {\bf 1})(+2)~,~~~
 N_i=4\times ({\bf 1},{\bf 1})(0)~.\nonumber
\end{eqnarray}
Here the $SU(3)_c\otimes SU(2)_w$ quantum numbers are given in bold font, 
whereas the $U(1)_Y$ hypercharge is given in parentheses. 
The {\em four} generations $(i=1,2,3,4)$ of quarks 
and leptons are given by $Q_i,D_i,U_i$ respectively $L_i,E_i,N_i$ 
(the chiral superfields $N_i$ correspond to the right-handed
neutrinos\footnote{As we will point out in the next section, Majorana
masses for neutrinos can be generated in the TSSM4 {\em without}
introducing {\em bulk} 
right-handed neutrinos but via a novel higher dimensional
mechanism.}).
The massive KK spectrum will be described in section IV where
we discuss the gauge coupling unification in the TSSM4.

{}In the first approximation we assume that there is no mixing (at least in
the lepton sector) between the
first three ($i=1,2,3$) and the fourth generations. In the next section we
will discuss how to achieve this using (discrete) gauge symmetries.
Then we identify the first three generations with the known quarks and
leptons, whereas the heavy fourth generation is new. The fourth generation
superfield $L_4$ is then identified with the electroweak Higgs superfield. 

{}Next, let $X$ be the ``spurion'' superfield whose F-term breaks
supersymmetry. The down-quark masses are
generated via the usual couplings in the superpotential which are of the
following type:  
\begin{equation}
 Q_i D_j L_4~,~~~i,j=1,2,3,4~.
\end{equation}
The up-quark masses, however, are generated via the following couplings in
the K\"ahler potential:
\begin{equation}
 Q_i U_j L_4^* X^*/M^2_s~,~~~i,j=1,2,3,4~.
\end{equation}
If the F-term $\langle F_X\rangle\sim M_s^2$, where the string scale 
$M_s$ is in the TeV
range\footnote{Here and in the following we assume that the string
coupling is of order one, so that the string scale $M_s$ and the
fundamental Planck scale $M_{Pf}$ are of the same order of
magnitude. Moreover, by ``TeV range'' we do not necessarily mean that $M_s$
is of order of a few TeV, but can be $\sim 10-100$ TeV. In fact, as was
pointed out in \cite{TSSM}, which we will re-iterate in the following, this
range for the string scale might be preferred from various phenomenological
considerations.}, then the up-quark Yukawa couplings are {\em not} suppressed.

{}The charged lepton masses for the first three generations are generated
via the usual couplings in the superpotential:
\begin{equation}
 L_a E_b L_4~,~~~a,b=1,2,3~.
\end{equation}
Note that the analogous couplings for the fourth generation such as 
$L_4 E_4 L_4$ are absent due to antisymmetry of the $SU(2)_w$
contractions. Thus, the desired Yukawa coupling for the fourth generation 
charged lepton must come from the K\"ahler potential. One possible coupling
of this type is the following:
\begin{equation}
 ({\cal D}X) X^* L_4({\cal D} L_4) E_4/M_s^4~.
\end{equation}
Once again, this coupling is not suppressed as long as 
$\langle F_X\rangle\sim M_s^2$. Generation of neutrino masses will be
discussed in the next section.

{}Finally, let us note that a light Higgs is no longer a prediction of
supersymmetry in the present context. Thus, in the TSSM4 the Higgs
potential can come from the following terms in the K\"ahler potential
(here $C$ is a model dependent numerical constant)
\begin{equation}
 (XX^*/M_s^2)(L_4 L_4^* +C(L_4 L_4^*)^2/M^2_s+\dots)~,
\end{equation}   
which can in principle tolerate any mass the Higgs could have in the
context of the non-supersymmetric Standard Model.

\section{Proton Stability and Neutrino Masses}

{}In this section we discuss proton stability and generation of small 
neutrino masses in the TSSM4. In particular, we consider gauging
$U(1)_{\cal B}$ and $U(1)_{\cal L}$ bulk symmetries (or, more precisely,
discrete subgroups thereof) corresponding to the baryon respectively
lepton numbers.

\subsection{Proton Stability} 

{}To ensure proton stability within the TSSM4, the simplest possibility is
to gauge the baryon number. Note that in the context of three-generation
models this is problematic as the corresponding $U(1)$ symmetry would be
anomalous. However, with four generations we can have an anomaly free
$U(1)$ gauge symmetry corresponding to the baryon number \cite{anto}. Indeed, 
consider the following $U(1)_{\cal B}$ charge assignments:
\begin{eqnarray}
 && Q_a:~+1~,~~~D_a:~-1~,~~~U_a:~-1~,~~~a=1,2,3~,\nonumber\\
 && Q_4:~-3~,~~~D_4:~+3~,~~~U_4:~+3~,\nonumber\\
 && L_i:~0~,~~~E_i:~0~,~~~N_i:~0~,~~~i=1,2,3,4~.\nonumber
\end{eqnarray}
Note that the above $U(1)_{\cal B}$ charge assignment is such that ${\cal
B}=3B$ for the first three generations, where $B$ is the usual baryon
number. The ${\cal B}$ charge assignment for the fourth generation quarks
is dictated by the anomaly cancellation requirement. Cancellation of the
$U(1)_{\cal B}$ anomalies can be seen by noting that the above $U(1)_{\cal
B}$ charge assignment can be viewed as descending from a ``vector-like''
$SU(4)$ ``flavor'' symmetry with the left handed quarks transforming in
${\bf 4}$ of $SU(4)$, and the left-handed anti-quarks transforming in
${\overline{\bf 4}}$. 

{}Note that the $U(1)_{\cal B}$ symmetry must somehow be broken or else it
would lead to a new long-range force contradicting the data. Here we have a
few different possibilities. First, we can break $U(1)_{\cal B}$, 
which in this case is a bulk gauge symmetry, on a distant brane,
so that this breaking is communicated to the brane on which
the Standard Model fields are localized via the large extra dimensional
bulk fields. The $U(1)_{\cal B}$ gauge symmetry breaking can then be (almost)
maximal on the distant brane, while on our brane it is suppressed by the
corresponding factor related to the volume of the large extra
dimensions. It is not difficult to see that the baryon number violating
operators are suppressed by the appropriate 
powers of $M_s/M_P$ (where $M_P$ is the four dimensional Planck scale)
provided that the separation between the branes is of order of the size of
the large extra dimensions. Moreover, the $U(1)_{\cal B}$ gauge boson picks
up a mass of order of or larger than an inverse millimeter leading to a new
force which could be observable in the sub-millimeter range \cite{Ant,TeV}.

{}Note, however, that if we break the $U(1)_{\cal B}$ gauge symmetry 
completely, then in certain cases there might be induced $\Delta B=1$ 
dimension five operators of the form
$QQQL/M_P$ without any additional suppression. Such operators
are disastrous for proton stability in the supersymmetric context (which we
adapt) via a one-loop graph involving a chargino exchange
\cite{wein}. Thus, 
to guarantee proton stability, instead of
breaking $U(1)_{\cal B}$ completely, we can consider breaking it to its
discrete subgroup which would forbid such dimension five
operators. Alternatively, we may completely bypass $U(1)_{\cal B}$ and
directly consider gauging such a discrete symmetry. In fact, 
a ${\bf Z}_2$
subgroup ${\widetilde {\cal B}}_2$ 
(in the following we will denote the ${\bf Z}_N$
subgroup of any $U(1)_{\cal A}$ gauge symmetry by ${\widetilde {\cal
A}}_N$) suffices for these purposes\footnote{Note that this ${\bf Z}_2$
discrete symmetry also forbids dimension four baryon number violating
operators of the form $UDD$.}. Moreover, if we gauge the ${\widetilde
{\cal B}}_2$ subgroup, then proton is completely stable.
Indeed, the higher dimensional
operators responsible for proton decay can be schematically written as 
$QQQL^{k+1}(L^*_4)^k$, and violate the baryon number by $\Delta B=1$, that
is, they violate the ${\cal B}$ number by $\Delta {\cal B}=3$. If, however, the
${\cal B}$ number is conserved modulo 2, which is the case when we gauge the
${\widetilde {\cal B}}_2$ discrete subgroup, then these operators are all
forbidden, so that proton is completely stable. 

{}Note, however, that $\Delta B=2$ processes would be allowed by the 
${\widetilde {\cal B}}_2$ symmetry. These must also be suppressed as there
are experimental bounds on such processes (for instance, from the
$NN\rightarrow \pi\pi$ transitions, where $N=p,n$). The corresponding
higher dimensional operators can be completely suppressed if instead of the
${\bf Z}_2$ subgroup ${\widetilde {\cal B}}_2$ of $U(1)_{\cal B}$ we gauge
its ${\bf Z}_4$ subgroup ${\widetilde {\cal B}}_4$. Indeed, $\Delta B=2$
implies $\Delta {\cal B}=6$, and since the ${\cal B}$ number is conserved
modulo 4 in this case, all such processes are forbidden. 

{}Finally, let us point out a ``hybrid'' possibility which might be of
interest in the light of the discussion in section III, where we consider
possible geometric embeddings of the TSSM4 in the brane world
context. Thus, consider gauging the ${\widetilde {\cal B}}_2$ discrete
symmetry with the $U(1)_{\cal B}\rightarrow {\widetilde {\cal B}}_2$
breaking taking place on a distant brane\footnote{Alternatively, we could
imagine that we start with the ${\widetilde {\cal B}}_4$ symmetry broken on
a distant brane to its ${\bf Z}_2$ subgroup ${\widetilde {\cal
B}}_2$.}. Then proton would be completely stable but the $\Delta B=2$
baryon number violating processes would be allowed albeit adequately
suppressed by the appropriate powers of $M_s/M_P$. 

{}The above ``hybrid'' possibility might
be interesting in the context of baryogenesis. It seems non-trivial to have
successful baryogenesis in the context of TeV-scale brane world for the
following reason. On the one hand, baryon number violating processes must
be adequately suppressed by various (discrete) gauge symmetries, and the
allowed baryon number operators then have rather high dimensions. On the
other hand, the reheat temperature cannot be too high in the context of
TeV-scale brane world or else one generically runs into various
cosmological problems (for instance, with the bulk gravitons)
\cite{TeV}. Here we would like to point out the following possibility. In
the above ``hybrid'' scenario, where ${\widetilde {\cal B}}_2$ is unbroken
but $U(1)_{\cal B}$ (or ${\widetilde {\cal B}}_4$) is broken down to
${\widetilde {\cal B}}_2$ on a distant brane, the dimension five operators
$QQQL$ are forbidden, but the operators of the form $QQQQQQL_4L_4$ (after
the electroweak breaking) give rise to effective six-fermion operators
(which violate the baryon number by $\Delta B=2$) suppressed by 
$\eta M_{\rm EW}^2/M_s^7$. Here the additional suppression factor $\eta$
(which is given by the appropriate powers of $M_s/M_P$) is due to the fact
that the above breaking occurs on a distant brane. However, in the early
universe the branes can be much closer to each other. In fact, even though
the sizes of the large extra dimensions can already be fixed at their zero
temperature values, the brane stabilization need not have taken place yet
as the latter may be due to dynamics different from that responsible for
radius stabilization. (For recent discussions on brane stabilization in
the brane world context, 
see, {\em e.g.}, \cite{dvali,branestab}.) Thus, when the
branes are at distances of order $1/M_s$ from each other, the factor
$\eta\sim 1$, so that there is no extra suppression in the above effective
six-fermion operators. We should point out that it is still not completely 
clear whether one can have successful baryogenesis in this context, and it
would be interesting to understand this issue in more detail. (For recent
discussions on baryogenesis in the context of TeV-scale brane world, see,
{\em e.g.}, \cite{benakli,savas,gabad}.)    

\subsection{Neutrino Masses}

{}To generate small (Dirac) neutrino masses in the TSSM4, we can assume
that the right-handed neutrinos $N_i$ are bulk fields, and then the
neutrino masses are generated via the higher dimensional mechanism of
\cite{neutrinos}. However, as was pointed out in \cite{proton}, for this
mechanism to work one must adequately suppress dangerous (effective) 
dimension five lepton number violating operators which in the TSSM4 context
have the following form: $L_a L_b L^*_4 L^*_4$, $a,b=1,2,3$. Here we take 
fermionic components for $L_a$ and $L_b$, and bosonic components for both 
$L^*_4$'s. If these operators are suppressed only by $1/M_s$, then upon the
electroweak symmetry breaking (that is, once the $L_4$ VEV is non-zero)
such operators would generate unacceptably large Majorana neutrino masses
of order $M_{\rm EW}^2/M_s$ for the first three generation neutrinos. On
the other hand, the analogous operator of the form $L_4 L_4 L_4^* L_4^*$
is welcome. In fact, the ``see-saw'' type of mass for the fourth
generation neutrino amounts to generating precisely such an operator via
the mixing of the fourth generation neutrino with the neutral gauginos (whose
linear combination in this case plays the role of the
corresponding ``right-handed'' neutrino). 

{}To suppress the operators of the form $L_a L_b L^*_4 L^*_4$, the simplest
possibility is to gauge the lepton number. Note that, just as in the case of
the baryon number, in the context of the three-generation models this is
problematic as the corresponding $U(1)$ symmetry would be
anomalous. However, with four generations we can have an anomaly free
$U(1)$ gauge symmetry corresponding to the lepton number. Indeed, consider
the following $U(1)_{\cal L}$ charge assignments:
\begin{eqnarray}
 && Q_i:~0,~~~D_i:~0~,~~~U_i:~0~,~~~i=1,2,3,4~,\nonumber\\
 && L_a:~+1~,~~~E_a:~-1~,~~~N_a:~-1~,~~~a=1,2,3~,\nonumber\\
 && L_4:~-3~,~~~E_4:~+3~,~~~N_4:~+3~.\nonumber
\end{eqnarray}
Note that the above  $U(1)_{\cal L}$ charge assignment is such that ${\cal
L}=L$ for the first three generations, where $L$ is the usual lepton
number. The ${\cal L}$ charge assignment for the fourth generation leptons is
dictated by the anomaly cancellation requirement. Cancellation of the
$U(1)_{\cal L}$ anomalies is completely analogous to that for $U(1)_{\cal B}$. 

{}Note that the $U(1)_{\cal L}$ gauge symmetry forbids the operators of the
form $L_a L_b L^*_4 L^*_4$ (but allows the operator $L_4 L_4 L^*_4
L^*_4$). Thus, gauging the $U(1)_{\cal L}$ symmetry would suffice for the
purposes of suppressing dangerous dimension five lepton number violating
operators. Note, however, that the $U(1)_{\cal L}$ symmetry is
unnecessarily strong for these purposes alone. Moreover, it actually
forbids the desirable (effective) Yukawa couplings of the form $QDL_4$,
$QUL_4^*$, $LEL_4$ as well as $LNL_4^*$ as $L_4$ carries a non-zero
$U(1)_{\cal L}$ charge. To remedy this, instead of gauging the full
$U(1)_{\cal L}$ symmetry, very much in the spirit of \cite{proton} 
we propose to gauge its ${\bf Z}_3$ subgroup
${\widetilde {\cal L}}_3$. The ${\widetilde {\cal L}}_3$ charge assignments
are given by:   
\begin{eqnarray}
 && Q_i:~0,~~~D_i:~0~,~~~U_i:~0~,~~~i=1,2,3,4~,\nonumber\\
 && L_a:~+1~,~~~E_a:~-1~,~~~N_a:~-1~,~~~a=1,2,3~,\nonumber\\
 && L_4:~0~,~~~E_4:~0~,~~~N_4:~0~.\nonumber
\end{eqnarray}
Note that the ${\widetilde {\cal L}}_3$ charge is conserved modulo 3. Thus,
the operators $L_a L_b L^*_4 L^*_4$ are still forbidden, while the
(effective) Yukawa couplings $QDL_4$,
$QUL_4^*$, $LEL_4$ as well as $LNL_4^*$ are now allowed.

{}Gauging the ${\widetilde {\cal L}}_3$ discrete symmetry allows for an
interesting novel higher dimensional
mechanism for generating small Majorana masses {\em
without} introducing {\em bulk} right-handed neutrinos. 
(A similar mechanism was pointed out in \cite{proton} in the TSSM context.)
Thus, note that the
mixing between the first three and the fourth generation leptons is absent
if the ${\widetilde {\cal L}}_3$ discrete symmetry is exact. However,
imagine that this symmetry is broken on a distant brane. Then generically 
there is going to be induced non-zero mixing between the first three and
the fourth generation leptons, but this mixing is suppressed by a factor of
$M_s/M_*$, where $M_*$ (which is roughly of order $M_P$, but can
actually be a few orders of magnitude smaller)
is related to the volume of the large extra
dimensions, and with the appropriate separation between the branes can be
of the correct order of magnitude so that the induced Majorana masses for
the first three generation neutrinos are in the desirable range. (Recall
that the fourth generation neutrino has a mass between 45 and 65 GeV.) Note
that this mechanism for generating small Majorana masses does {\em not} 
require {\em bulk} right-handed neutrinos. The role of the {\em fourth}
generation ``right-handed'' neutrino is played by a linear combination of
neutral gauginos, and the first three generations acquire Majorana masses
via the mixing with the fourth generation neutrino. Nonetheless, the
presence of the states $N_i$ in this scenario is still required by anomaly
cancellation, but what is different now is that these states need not be
bulk fields. Thus, we can assume that they do {\em not} propagate in the
large extra dimensions. However, if they are localized on the same brane as
the fields $L_i$, then we would generically 
have unacceptably large Dirac neutrino
masses coming from the allowed couplings $LNL_4^*$. This can be remedied by
assuming that the states $N_i$ are localized on a distant brane (or,
alternatively, they come from different fixed points in the context of 
orbifold compactifications\footnote{Thus, for instance, if we assume that
our brane is sitting at an orbifold fixed point while the fields $N_i$ come
from a {\em different} fixed point, then the couplings $LNL_4^*$ would be
suppressed by an exponential factor $\sim \exp(-c(RM_s)^2)$, where $R$ is
(roughly) the linear ``size'' of dimensions along which the orbifold fixed 
points are
separated (and these dimensions are assumed to be large enough so that this
exponential factor is adequately suppressed), while $c$ is a model dependent
numerical constant of order 1.} - 
see the next section), in which case the above
couplings are at least exponentially suppressed.  

\section{Gauge Coupling Unification}

{}In this section we discuss the gauge coupling unification in TSSM4. Since
the unification scale, which we identify with the string scale, is now in
the TeV range, the unification mechanism must be very different from that
in the MSSM. In \cite{dienes} such a unification mechanism was
proposed. Thus, let the Standard Model fields be localized on D$p$-branes
(where $p=4,5$) with $p-3$ extra directions compactified on a space with
linear ``size(s)'' of order $R$. Let us assume that $R\gg 1/M_s$. Then
between the Kaluza-Klein (KK) threshold $1/R$ and the cut-off $M_s$ we have
$\sim (RM_s)^{p-3}$ heavy KK modes carrying the corresponding gauge quantum
numbers. These states contribute into the gauge coupling running above the
scale $1/R$, and if we are interested in the contributions of these states
(that is the corresponding thresholds)
into the low energy (that is, at scales $\mu \ll 1/R\ll M_s$) gauge
coupling renormalization at one loop, they are proportional to the number
of these states $N\sim (RM_s)^{p-3}\gg 1$. Thus, one can hope that there
might be a possibility of an {\em accelerated} unification at $M_s$ in the
TeV range.

{}As was explained in detail in \cite{TSSM,Taylor}, this mechanism would
have no predictive power unless we adapt the supersymmetric context where
the zero modes of the compactification (corresponding, in particular, to the
Standard Model fields) are ${\cal N}=1$ supersymmetric, while the heavy KK
modes have ${\cal N}=2$ supersymmetry. The reason is that higher loop
effects would otherwise be as large as the one-loop threshold contribution
as we have a large number $N$ of heavy KK modes circulating in loops. The
underlying ${\cal N}=2$ supersymmetry at the heavy KK modes guarantees, as
was shown in \cite{TSSM,Taylor}, that the higher loop effects are
subleading compared with the leading one-loop threshold contribution. This
is due to the well known non-renormalization property in ${\cal N}=2$ gauge
theories. Note, however, that, as was stressed in \cite{TSSM,Taylor}, higher
loop cancellations due to the underlying supersymmetry persist after
supersymmetry breaking only if $M_{\rm SUSY}\ll 1/R$. This is one of the 
key reasons why we assume that $M_s$ is at least $10-100$ TeV.  

{}We will discuss possible geometric embeddings (via orbifolds) of such
compactifications in the next subsection. Here let us just assume that the
zero modes are ${\cal N}=1$ supersymmetric with the one-loop
$\beta$-function coefficients $b_r$, $r=1,2,3$ (here $r$ labels the
corresponding subgroup of $SU(3)_c\otimes SU(2)_w \otimes U(1)_Y$), while
the heavy KK levels are populated by ${\cal N}=2$ supermultiplets with the     
one-loop $\beta$-function coefficients ${\widetilde b}_r$. It is then not
difficult to show that the unification of gauge couplings in such a model is
as precise as in the MSSM (at one loop) if and only if the following
constraint is satisfied:
\begin{equation}\label{nu}
 \mbox{$\nu_{rs}\equiv\nu$ is independent of $r,s$, where}~~
 \nu_{rs}\equiv{{\widetilde b}_r-{\widetilde b}_s\over b^*_r-b^*_s}~~
 \mbox{for
 $r\not=s$.}  
\end{equation}
Here $b^*_r$ are the one-loop $\beta$-function coefficients in the MSSM:
$b^*_1=33/5$, $b^*_2=1$, $b^*_3=-3$ (we are using the standard
normalization for the $U(1)_Y$ gauge coupling $\alpha_1=(5/3)\alpha_Y$). 

{}In the following we assume that at the heavy KK levels we have ${\cal
N}=2$ vector multiplets in the adjoint of the gauge group, and the charged 
${\cal N}=2$ hypermultiplets can only have the gauge quantum numbers of
$Q,D,U,L,E$ (that is, we have no ``exotic'' matter). Moreover, we will not
distinguish between the $Q,D,U,L,E$ and their conjugate quantum numbers as
the difference is immaterial as far as the one-loop gauge coupling running
is concerned. We will denote the number of the corresponding charged
hypermultiplets (per heavy KK mass level) by $n_Q,n_D,n_U,n_L,n_E$. 

{}The one-loop $\beta$-function coefficients ${\widetilde b}_r$ are given
by:
\begin{eqnarray}  
 &&{\widetilde b}_1={1\over 5}\left( n_Q + 2n_D + 8n_U + 3n_L +
 6n_E\right)~,\nonumber\\ 
 &&{\widetilde b}_2=-4+ 3n_Q + n_L~,\nonumber\\
 &&{\widetilde b}_3=-6+ 2n_Q + n_D + n_U~.\nonumber 
\end{eqnarray}
The constraint (\ref{nu}) can then be rewritten as 
\begin{eqnarray} 
 &&n_Q+n_L-n_D-n_U=4\nu-2~,\nonumber\\
 &&n_U+n_E-2n_Q=6\nu-4~.\nonumber
\end{eqnarray}
It is straightforward to analyze this system of algebraic equations, but,
instead of being most general here, for illustrative purposes let us
consider possible solutions with $\nu=1$. One such solution is
$n_Q=n_D=n_U=n_f$, $n_L=n_E=n_f+2$. Note that this precisely corresponds to
the TSSM matter content of \cite{TSSM} with $n_f$ generations plus $H_\pm$
and $F_\pm$ hypermultiplets (see \cite{TSSM} for details) propagating
in the $p-3$ dimensional bulk (where the gauge bosons live).    

{}Here we are interested in a four-generation model (that is, the TSSM4),
where we  do {\em not} wish to add any extra ``vector-like'' states (such as
$H_\pm$ of the MSSM, or $F_\pm$ of the TSSM). One interesting (for the
reason we will explain in a moment) solution of this type is the following.
Let $n_Q=n_D=n_E=0$, $n_U=2$, and $n_L=4$. Then we can imagine the following
setup. We have four generations of quarks and leptons only. The states
$Q_i,D_i,E_i$, $i=1,2,3,4$, are localized at fixed points (see the next
subsection), so that they do not propagate in the $p-3$ dimensional bulk.  
The same is the case for the states $U_1,U_2$. The states $U_3,U_4$ as well
as $L_i$, $i=1,2,3,4$, however, are {\em not} localized at fixed points,
but propagate in the $p-3$ dimensional bulk. Then the unification in the
TSSM4 within this setup is as precise at one loop as in the MSSM and the TSSM.

\subsection{Brane World Embedding}

{}As we mentioned above, the massless modes in the TSSM4 are ${\cal N}=1$
supersymmetric, whereas the massive KK modes have ${\cal N}=2$
supersymmetry. Such spectra arise in orbifold compactifications in the
brane world context, more concretely, in compactifications on generalized
Voisin-Borcea orbifolds. Thus, consider Type I/Type I$^\prime$ string
theory compactified on an elliptically fibered 
Calabi-Yau three-fold (with $SU(3)$ holonomy) of
the following form: ${\cal M}_3=(T^2\otimes {\mbox{K3}})/{\bf Z}_M$, where
the generator $g$ of ${\bf Z}_M$ (here $M$ can only take values $M=2,3,4,6$
as the action of $g$ on $T^2$ must be crystallographic) acts as a $2\pi/M$ 
rotation $g z_1 =\omega z_1$ on $T^2$ ($z_1$ is the complex coordinate
parametrizing $T^2$, and $\omega\equiv\exp(2\pi i/M)$), 
and as $g\Omega_2=\omega^{-1} \Omega_2$ on K3
($\Omega_2$ is the holomorphic 2-form on K3). Next, consider D5-branes
wrapping the fiber $T^2$. Then the zero modes of the gauge theory living in
the world-volume of the D5-branes are ${\cal N}=1$ supersymmetric, while
the heavy KK modes are ${\cal N}=2$ supersymmetric.

{}In the following we would like to discuss possible brane world embeddings 
of the TSSM4 in the context of such orbifold compactifications. In
particular, solutions to the constraint (\ref{nu}) imply that some of the
matter fields should be localized at fixed points while others propagate in the
entire world-volume of D5-branes. It is therefore necessary to check
whether any such solution is at least in principle compatible with the
geometric embedding via orbifolds\footnote{This was checked for the TSSM in
\cite{TSSM,proton}. Subsequently other straightforward variations of the TSSM
were discussed in the literature (see, {\em e.g.},\cite{other}),
where some fields are
supposed to be localized at fixed points while others propagate in the
bulk. However, it is not enough to simply find a solution to the constraint
(\ref{nu}), which is straightforward, but one must also check whether such
a solution is at least 
conceivably embeddable in the brane world context.}. Here we
should stress that the following discussion still does not guarantee that
the corresponding model exists as due to the lack of necessary model
building technology it is not clear at present whether there exists the
appropriate choice of the gauge bundle which would give the desired gauge
group and spectrum\footnote{Here we would like to point out that, 
if such an embedding exists, the string coupling is expected to be of order 1
as otherwise it is difficult to imagine how the dilaton could be stabilized 
\cite{BW}. The weakness of the Standard Model gauge couplings then requires 
that the volume of the compact dimensions inside of the D5-branes on which the
Standard Model gauge fields are localized be somewhat larger than 1 
(in the string units) as is the case in the context of unification via 
Kaluza-Klein thresholds. Note that even though the underlying string theory is
in the non-perturbative regime, the corresponding effective field theory is 
weakly coupled, so that various perturbative considerations such as gauge 
coupling unification are valid. In fact, in the present context this issue
was discussed in detail in \cite{TSSM,Taylor}, where the importance of 
the underlying ${\cal N}=2$ supersymmetry at the heavy Kaluza-Klein levels 
was stressed.}. However, as we discuss in the following, 
the purely geometric part of this issue
can be studied by considering possible orbifold compactifications with the
corresponding discrete gauge symmetries.           

{}Our discussion in this subsection will be brief as most of the
ingredients we are going to use here were discussed in detail in
\cite{TSSM}. Thus, we wish to obtain a model where the fields
$Q_i,D_i,E_i,U_{1,2}$ ($i=1,2,3,4$) 
are localized at fixed points, while the fields $L_i,U_{3,4}$ propagate in
the bulk (see above). As was pointed out in \cite{TSSM}, the fields
localized at the orbifold fixed points arise in the Type I/Type I$^\prime$
compactifications in the context of {\em non-perturbative} orientifolds
examples of which were recently constructed in \cite{np} (for recent
progress in perturbative orientifolds, see, {\em e.g.}, \cite{TypeI}). In
particular, these states arise in the twisted sectors, and, therefore,
carry the corresponding discrete quantum numbers. Thus, we will first gauge
an anomaly free discrete symmetry under which the states
$Q_i,D_i,U_i,L_i,E_i,N_i$ have certain discrete gauge charge assignments, and
then we will identify this discrete symmetry with that of an orbifold
compactification. This will suggest a possible brane world embedding of the
TSSM4 with the above matter content.

{}Our strategy for gauging the desired discrete symmetry will be to first   
gauge an anomaly-free continuous Abelian symmetry, and then to restrict to
its discrete subgroup. Thus, consider the $U(1)_{\cal Z}$ symmetry, where
${\cal Z}\equiv 3(Y-{\cal R})$, and the ${\cal Z}$ action is the same on all
four generations (that is, the $U(1)_{\cal Z}$ symmetry is generation
blind). Here $Y$ is the hypercharge, while the ${\cal R}$ charge assignment
(which is generation blind) is given by:
\begin{eqnarray}
 &&{\cal R}_Q=0~,~~~{\cal R}_D=+1~,~~~{\cal R}_U=-1~,\nonumber\\
 &&{\cal R}_L=0~,~~~{\cal R}_E=+1~,~~~{\cal R}_N=-1~.\nonumber
\end{eqnarray}  
Note that the $U(1)_{\cal R}$ symmetry is anomaly free and compatible with
the $U_{\cal B}$ and $U_{\cal L}$ symmetries. This implies that so is
$U(1)_{\cal Z}$. In fact, the ${\cal Z}$ charge assignment is such that
${\cal Z}$ acts as $3(B-L)$ on the first three generations.

{}In the following we will be interested in gauging not $U(1)_{\cal Z}$ but
rather $U(1)_{{\cal Z}^\prime}$, where ${\cal Z}^\prime\equiv {\cal
Z}-{\cal R}$. The $U(1)_{{\cal Z}^\prime}$ charge assignment is given by:
\begin{eqnarray}
 &&{\cal Z}^\prime_Q=+1~,~~~{\cal Z}^\prime_D=-2~,~~~
 {\cal Z}^\prime_U=0~,\nonumber\\
 &&{\cal Z}^\prime_L=-3~,~~~{\cal Z}^\prime_E=+2~,~~~{\cal Z}^\prime_N=+4~.
 \nonumber
\end{eqnarray}  
Let ${\widetilde {\cal Z}}^\prime_3$ be a ${\bf Z}_3$ subgroup of
$U(1)_{{\cal Z}^\prime}$, and let $\theta$ be the generator of 
${\widetilde {\cal Z}}^\prime_3$. The ${\widetilde {\cal Z}}^\prime_3$
charge assignment is given by (the ${\widetilde {\cal Z}}^\prime_3$ charge
is conserved modulo 3):
\begin{eqnarray}
 &&Q_i:~+1~,~~~D_:~+1~,~~~U_i:~0~,\nonumber\\
 &&L_i:~0~,~~~E_i:~-1~,~~~N_i:~+1~.\nonumber
\end{eqnarray}
Here $i=1,2,3,4$, and ${\widetilde {\cal Z}}^\prime_3$ is a generation
blind symmetry.

{}Next, consider the $U(1)_{{\cal R}^\prime}$ symmetry, where ${\cal
R}^\prime$ acts as ${\cal R}$ on the first two generations, while the other
two generations have zero ${\cal R}^\prime$ charges. This symmetry is
anomaly free and compatible with all the other symmetries we have
considered so far. In the following we will need a ${\bf Z}_2$ subgroup 
${\widetilde {\cal R}}^\prime_2$ of  
$U(1)_{{\cal R}^\prime}$, whose generator we will denote by $R$.

{}The discrete symmetry we are interested in here is the ${\bf Z}_6$
symmetry ${\widetilde {\cal Z}}^\prime_3\otimes 
{\widetilde {\cal R}}^\prime_2$. In particular, we will identify the
generator $\theta R$ of this group with the generator $g$ of the orbifold
group $Z_M$ with $M=6$ in the above discussion. Then the states
$L_i,U_{3,4}$ come from the untwisted sector, the states
$Q_i,D_{3,4},N_{3,4}$ come from the $\theta$-twisted (that is,
$g^{-2}$-twisted) sector, the states $E_{3,4}$ come from the
$\theta^{-1}$-twisted (that is, $g^2$-twisted) sector, the states
$D_{1,2},U_{1,2},N_{1,2}$ come from the $\theta R$-twisted (that is,
$g$-twisted) sector, and the states $E_{1,2}$ come from the $\theta^{-1}
R$-twisted (that is, $g^{-1}$-twisted) sector. The $R$-twisted (that is,
$g^3$-twisted sector) does not contain any states. With the above
identifications, it is conceivable that the above model may be obtainable
as a Type I/Type I$^\prime$ compactification on a generalized Voisin-Borcea
orbifold of the form $(T^2\otimes {\mbox{K3}})/{\bf Z}_6$. Whether the
number of the corresponding fields in each sector (as well as their gauge
quantum numbers) can come out right is a model building question which
depends on an appropriate choice of the gauge bundle (whose existence is a
non-trivial issue) as discussed in detail in \cite{TSSM}.

{}Before we end this section, let us point out that the ${\widetilde {\cal
L}}_3$ and ${\widetilde {\cal B}}_2$ discrete gauge symmetries may also be
embeddable in the above context if we take K3 to be an orbifold K3 with the
orbifold group being ${\bf Z}_6^\prime\approx 
{\widetilde {\cal L}}_3 \otimes {\widetilde {\cal B}}_2$, where the
generator $g^\prime$ of ${\bf Z}_6^\prime$ acts trivially on the fiber 
$T^2$ . Note that we
could not embed ${\widetilde {\cal B}}_4$ instead of ${\widetilde {\cal
B}}_2$ in this context as the total orbifold group in this case could not
act crystallographically. This is the reason why we considered a ``hybrid''
possibility for suppressing baryon number violation in subsection III.A.

{}Also note that the above 
${\bf Z}_6$ discrete symmetry must actually be broken
or else the desirable Yukawa couplings of the Standard Model fermions to
the Higgs would be forbidden. This implies that we are actually talking
about (at least partially) {\em blown-up} orbifolds. In fact, if the
breaking of ${\bf Z}_6$ can be viewed as being due to some fields charged under
${\bf Z}_6$ acquiring non-zero VEVs in an asymmetric fashion (that is, the
VEV of the field with the $+1$ ${\bf Z}_6$ charge is different from that of
the field with the $-1$ ${\bf Z}_6$ charge, which is conceivable in the
context of almost completely broken supersymmetry), then this could be used to
explain the {\em vertical} hierarchy in the quark and lepton sectors.

{}Finally, we note that if we consider the ${\bf Z}_6\otimes {\bf
Z}_6^\prime$ orbifold, then the right-handed neutrinos can be seen to be
localized at ``(3+1)-dimensional'' fixed points, that is, they do not
propagate in large extra dimensions in this case. This is one of the
reasons why the novel
mechanism for generating small Majorana neutrino masses discussed in
subsection III.B might be appealing in this context. 

\section{Other Issues}

{}In this section we would like to briefly discuss some other issues in the
TSSM4 such as adequate suppression of FCNCs, and also a novel possibility
for supersymmetry breaking.

{}Suppressing (effective) four-fermion operators potentially inducing
unacceptably large FCNCs in the TSSM was discussed in detail in
\cite{flavor}. There it was shown that a non-Abelian discrete flavor group
$T_L\otimes T_R$ (where $T$ is the tetrahedral subgroup of $SU(2)$)
accompanied by a vector-like $U(1)_V$ flavor symmetry (or its appropriate
discrete subgroup) can adequately suppress all such operators. Here
the $T_L\otimes T_R$ discrete flavor symmetry acts only on the first
two generation quarks and leptons, whereas the $U(1)_V$ flavor symmetry acts
only on the third generation quarks and leptons (in the flavor basis). The
former suppresses four-fermion operators involving only 
the first two generation
states, while the latter guarantees that four-fermion operators also 
involving the third generation states are suppressed as well. 

{}The above mechanism for suppressing FCNCs can also be applied to the
TSSM4. The only difference is that we now have the fourth generation,
and we must also worry about possible FCNCs coming from the mixing between
the first three and the fourth generations. Note, however, that this mixing
is guaranteed to be small in the quark sector if the $U(1)_{\cal
B}\rightarrow {\widetilde {\cal B}}_2$ breaking occurs on a distant brane
as in the ``hybrid'' scenario discussed in subsection III.A. As to the
lepton sector, in the limit of unbroken ${\widetilde {\cal L}}_3$ discrete
gauge symmetry the mixing between the first three and the fourth generation
leptons is absent, and, since this discrete gauge symmetry is broken on a 
distant brane, the resulting FCNCs are completely adequately suppressed. 

{}Finally, we would like to point out that in the TSSM4 there {\em a
priori} exists a novel possibility for supersymmetry breaking. In
particular, imagine that not only the lowest (that is, scalar) component of
the chiral superfield $L_4$ has a non-zero VEV, but also the highest
component (that is, the F-term) develops a non-zero VEV. Then the F-term
$F_{L_4}$ could be responsible for supersymmetry breaking. In fact, in this
case we might no longer need a ``spurion'' superfield $X$ introduced in section
II. Thus, the up-quark masses could be generated from the following
dimension five operators in the K\"ahler potential: $QUL_4^*/M_s$. Indeed,
if $\langle F_{L_4} \rangle\sim M_s^2$, then we have effective Yukawa
couplings in the up-quark sector. Similar consideration apply, for
instance, to the Yukawa couplings for the fourth generation charged
leptons. It would be interesting to investigate this new possibility for
supersymmetry breaking in the TSSM4 in more detail.

\acknowledgments

{}I would like to thank Gregory Gabadadze and Aaron Grant for discussions. 
This work was supported in part by the grant NSF PHY-96-02074, and
the DOE 1994 OJI award. I
would also like to thank Albert and Ribena Yu for financial support.


\begin{references}

\bibitem{TeV} N. Arkani-Hamed, S. Dimopoulos and G. Dvali, 
Phys. Lett. {\bf B429} (1998) 263; Phys. Rev. {\bf D59} (1999) 0860004.

\bibitem{GK} A.K. Grant and Z. Kakushadze, hep-ph/9906556.

\bibitem{TSSM} Z. Kakushadze, Nucl. Phys. {\bf B548} (1999) 205.

\bibitem{dienes} K.R. Dienes, E. Dudas and T. Gherghetta, Phys. Lett. 
{\bf B436} (1998) 55; Nucl. Phys. {\bf B537} (1999) 47; hep-ph/9807522.

\bibitem{Ant} I. Antoniadis, Phys. Lett. {\bf B246} (1990) 317.

\bibitem{TV} T.R. Taylor and G. Veneziano, Phys. Lett. {\bf B212} (1988)
147. 

\bibitem{Taylor} Z. Kakushadze and T.R. Taylor, hep-th/9905137.

\bibitem{proton} Z. Kakushadze, Nucl. Phys. {\bf B552} (1999) 3.

\bibitem{flavor} Z. Kakushadze, Nucl. Phys. {\bf B551} (1999) 549.

\bibitem{anto} I. Antoniadis, N. Arkani-Hamed, S. Dimopoulos and G. Dvali,
Phys. Lett. {\bf B436} (1998) 257.

\bibitem{wein} S. Weinberg, Phys. Rev. {\bf D26} (1982) 287;\\
N. Sakai and T. Yanagida, Nucl. Phys. {\bf B197} (1982) 533.

\bibitem{ST} G. Shiu and S.-H.H. Tye, Phys. Rev. {\bf D58} (1998) 106007.

\bibitem{BW} Z. Kakushadze and S.-H.H. Tye, Nucl. Phys. {\bf B548} (1999) 180.

\bibitem{dvali} G. Dvali, hep-ph/9905204.

\bibitem{branestab} Z. Kakushadze, hep-th/9906222.

\bibitem{benakli} K. Benakli and S. Davidson, Phys. Rev. {\bf D60}
(1999) 025004.

\bibitem{savas} N. Arkani-Hamed, S. Dimopoulos, N. Kaloper and
J. March-Russell, hep-ph/9903224.

\bibitem{gabad} G. Dvali and G. Gabadadze, hep-ph/9904221.

\bibitem{neutrinos} N. Arkani-Named, S. Dimopoulos, G. Dvali and
J. March-Russell, talk presented by S. Dimopoulos at SUSY98, hep-ph/9811448.

\bibitem{other} C.D. Carone, Phys. Lett. {\bf B454} (1999) 70;\\
A. Delgado and M. Quir{\'o}s, hep-ph/9903400;\\
P.H. Frampton and A. Ra{\v s}in, hep-ph/9903479.

\bibitem{np} Z. Kakushadze, Phys. Lett. {\bf B455} (1999) 120; 
hep-th/9904211; hep-th/9905033.

\bibitem{TypeI} M. Berkooz and R.G. Leigh, Nucl. Phys. {\bf B483} (1997) 187;\\
C. Angelantonj, M. Bianchi, G. Pradisi, A. Sagnotti and 
Ya.S. Stanev, Phys. Lett. {\bf B385} (1996) 96;\\
Z. Kakushadze, Nucl. Phys. {\bf B512} (1998) 221; 
Nucl. Phys. {\bf B529} (1998) 157; Phys. Lett. {\bf B434} (1998) 269;
Nucl. Phys. {\bf B535} (1998) 311; Phys. Rev. {\bf D58} (1998) 101901;\\
Z. Kakushadze and G. Shiu, Phys. Rev. {\bf D56} (1997) 3686; 
Nucl. Phys. {\bf B520} (1998) 75;\\
G. Zwart, Nucl. Phys. {\bf B526} (1998) 378;\\
G. Aldazabal, A. Font, L.E. Ib{\'a}{\~n}ez and G. Violero,
Nucl. Phys. {\bf B536} (1998) 29;\\
Z. Kakushadze, G. Shiu and S.-H.H. Tye, 
Nucl. Phys. {\bf B533} (1998) 25;\\
J. Lykken, E. Poppitz and S.P. Trivedi, Nucl. Phys. {\bf
B543} (1999) 105;\\
R. Blumenhagen and A. Wisskirchen, Phys. Lett. {\bf B438}
(1998) 52;\\
Z. Kakushadze and S.-H.H. Tye, Phys. Rev. {\bf D58} (1998) 126001.



\end{references}
\end{document}